\documentclass[article]{mesa-NS}
\usepackage{times,mathptm}
\usepackage{amsmath,amsfonts,amssymb}
\usepackage{graphicx,graphics,epsfig}
\usepackage{fancyhdr,fancybox}
\usepackage{verbatim}
\usepackage{color}

\RequirePackage[T1]{fontenc}
\RequirePackage{epsf}
\RequirePackage{psfig}
\RequirePackage{mathptmx}      
\RequirePackage{flushend}
\RequirePackage[numbers]{natbib}
\RequirePackage[colorlinks,citecolor=blue,urlcolor=blue,linkcolor=blue]{hyperref}

\jvol{22}
\jnum{3}
\jyear{2015}

\numberwithin{equation}{section}
\numberwithin{theorem}{section}
\numberwithin{definition}{section}
\numberwithin{lemma}{section}
\numberwithin{corollary}{section}
\numberwithin{proposition}{section}
\numberwithin{example}{section}
\numberwithin{remark}{section}

\begin{document}
\label{pageinit}

\date{}

\title{Escape dynamics in a Hamiltonian system with four exit channels}

\author{Euaggelos E. Zotos $^{\star}$}

\maketitle

\noindent $ $ Department of Physics, Aristotle University of Thessaloniki,\\
           GR-541 24, Thessaloniki, Greece\\

\noindent $^\star$ \emph{Corresponding Author}. E-mail address: evzotos@physics.auth.gr

\markboth{}{Escape dynamics in a Hamiltonian system with four exit channels}

\footnotetext[2010]{\textit{{\bf Mathematics Subject Classification}}: Primary:  }

\begin{abstract}
We reveal the escape mechanism of orbits in a Hamiltonian system with four exit channels composed of two-dimensional perturbed harmonic oscillators. We distinguish between trapped chaotic, non-escaping regular and escaping orbits by conducting a thorough and systematic numerical investigation in both the configuration and the phase space. We locate the different basins of escape and we relate them withe the corresponding escape times of orbits. The SALI method is used for determining the ordered or chaotic nature of the orbits. It was observed that trapped and non-escaping orbits coexist with several escape basins. When the energy is very close to the escape energy the escape rate of orbits is huge, while as the value of the energy increases the orbits escape more quickly to infinity. Furthermore, initial conditions of orbits located near the boundaries of the basins of escape and also in the vicinity of the fractal domains were found to posses the largest escape rates. The degree of the fractality of the phase space was calculated as a function of the value of the energy. Our results were compared with earlier related work.
\end{abstract}

\section{Introduction}
\label{intro}

One of the most well studied subjects in nonlinear dynamics is the issue of escaping particles from dynamical systems. In particular, the issue of escapes in Hamiltonian systems is directly related to the subject of chaotic scattering which has been a very active field of research over the last decades (e.g., [\citealp{BTS96} -- \citealp{BGOB88}, \citealp{CPR75}, \citealp{C90}, \citealp{CK92}, \citealp{E88}, \citealp{JS88}, \citealp{ML02} -- \citealp{PH86}, \citealp{SASL06} -- \citealp{SS10}]). There are some types of Hamiltonian systems which have a finite energy of escape. This means that for values of energy lower than the escape energy the equipotential surfaces are closed thus making escape impossible. For values of energy larger than the escape energy on the other hand, these surfaces are open and several exit channels emerge through which the particles are now free to escape to infinity. In fact the literature is replete with studies of open Hamiltonian systems (e.g., [\citealp{BBS09}, \citealp{CE04}, \citealp{CKK93}, \citealp{CHLG12}, \citealp{EJSP08}, \citealp{EP14}, \citealp{KSCD99}, \citealp{NH01}, \citealp{STN02}, \citealp{SCK95} -- \citealp{SKCD96}, \citealp{Z15a}]).

In general terms, the infinity acts as an attractor for an escape particle, which may escape through different channels (exits) on the equipotential curve or on the equipotential surface depending whether the dynamical system has two or three degrees of freedom, respectively. Therefore, it is quite possible to obtain basins of escape, similar to basins of attraction in dissipative systems or even the Newton-Raphson fractal structures. Basins of escape have been studied in several papers (e.g., [\citealp{BGOB88}, \citealp{C02}, \citealp{KY91}, \citealp{PCOG96}]). The reader can find more details regarding basins of escape in [\citealp{C02}].

Without any doubt, the most characteristic model for time-independent open Hamiltonian systems of two degrees of freedom is the H\'{e}non-Heiles system [\citealp{HH64}]. Over the years an extensive load of research has been devoted on the escape properties of this system (e.g., [\citealp{AVS01} -- \citealp{AVS03}, \citealp{BBS08}, \citealp{BSBS12}, \citealp{dML99}, \citealp{S07}, \citealp{Z15b}]). Here it should be emphasized that all the above-mentioned references on escapes in the H\'{e}non-Heiles Hamiltonian system are exemplary rather than exhaustive, taking into account that a vast quantity of related literature exists.

Dynamical systems composed of perturbed harmonic oscillators has been extensively applied for describing local motion (i.e., near an equilibrium point) (e.g., [\citealp{AEFR06}, \citealp{C93}, \citealp{CK98} -- \citealp{CZ12}, \citealp{FLP98a} -- \citealp{HH64}, \citealp{SI79}, \citealp{Z12a} -- \citealp{Z14}]). Researchers have used either numerical (e.g., [\citealp{CZ12}, \citealp{KV08}, \citealp{ZC12}]) or analytical methods (e.g., [\citealp{CB82}, \citealp{D91}, \citealp{DE91}, \citealp{E00}, \citealp{ED99}]) in an attempt to explore and comprehend the orbital properties in these systems. Thus, taking into account all the above-mentioned facts, we decided to use a potential of a perturbed harmonic oscillator which has four escape channels in the configuration $(x,y)$ space (e.g., [\citealp{C90}, \citealp{CK92}, \citealp{CKK93}, \citealp{KSCD99}, \citealp{NH01}]). The aim of this work, is twofold: (i) to distinguish between trapped, escaping and non-escaping orbits and (ii) to locate the basins of escape leading to different escape channels and try to connect them with the corresponding escape times of the orbits.

The paper is organized as follows: in Section \ref{modpot} we present in detail the properties of the Hamiltonian system with four escape channels. All the computational techniques used in order to determine the character (ordered versus chaotic and trapped versus escaping) of the orbits are described in Section \ref{cometh}. In the following Section we conduct a thorough and systematic numerical analysis of several sets of initial conditions of orbits in both the configuration and the phase space. Our paper ends with Section \ref{disc}, where the main conclusions this work are discussed.

\section{Properties of the Hamiltonian system}
\label{modpot}

The general potential function of a two-dimensional perturbed harmonic oscillator is
\begin{equation}
V(x,y) = \frac{1}{2}\left(\omega_1^2 x^2 + \omega_2^2 y^2 \right) + \varepsilon V_1(x,y),
\label{genform}
\end{equation}
where $\omega_1$ and $\omega_2$ are the unperturbed frequencies of oscillations along the $x$ and $y$ axes, respectively, $\varepsilon$ is the perturbation parameter, while $V_1$ is the function containing the perturbing terms.

The function with the perturbation terms $(V_1(x,y))$ plays a key role as it determines the location as well as the total number of the escape channels in the configuration $(x,y)$ space. In our case where the channels of escape are four we have
\begin{equation}
V(x,y) = \frac{1}{2}\left(x^2 + y^2 \right) - x^2 y^2.
\label{pot}
\end{equation}
Without the loss of generality we consider the case where the perturbed harmonic oscillator is at the 1:1 resonance with common frequency $\omega_1 = \omega_2 = 1$, while for more convenient calculations we take $\varepsilon = -1$.

Then the Hamiltonian to potential (\ref{pot}) reads
\begin{equation}
H = \frac{1}{2}\left(\dot{x}^2 + \dot{y}^2 + x^2 + y^2\right) - x^2 y^2 = h,
\label{ham}
\end{equation}
where $\dot{x}$ and $\dot{y}$ are the momenta per unit mass conjugate to $x$ and $y$, respectively, while $h > 0$ is the numerical value of the Hamiltonian, which is conserved. We observe that the Hamiltonian $H$ is invariant under $x \rightarrow - x$ and/or $y \rightarrow - y$.

Consequently the equations of motion for a test particle with a unit mass $(m = 1)$ are
\begin{equation}
\ddot{x} = - \frac{\partial V}{\partial x}, \ \ \
\ddot{y} = - \frac{\partial V}{\partial y},
\label{eqmot}
\end{equation}
where, as usual, the dot indicates derivative with respect to the time. Furthermore, the variational equations governing the time evolution of a deviation vector $\vec{w} = (\delta x, \delta y, \delta \dot{x}, \delta \dot{y})$ are
\begin{eqnarray}
\dot{(\delta x)} &=& \delta \dot{x}, \nonumber \\
\dot{(\delta y)} &=& \delta \dot{y}, \nonumber \\
(\dot{\delta \dot{x}}) &=& -\frac{\partial^2 V}{\partial x^2}\delta x - \frac{\partial^2 V}{\partial x \partial y}\delta y, \nonumber \\
(\dot{\delta \dot{y}}) &=& -\frac{\partial^2 V}{\partial y \partial x}\delta x - \frac{\partial^2 V}{\partial y^2}\delta y.
\label{variac}
\end{eqnarray}

\begin{figure*}[!tH]
\centering
\resizebox{\hsize}{!}{\includegraphics{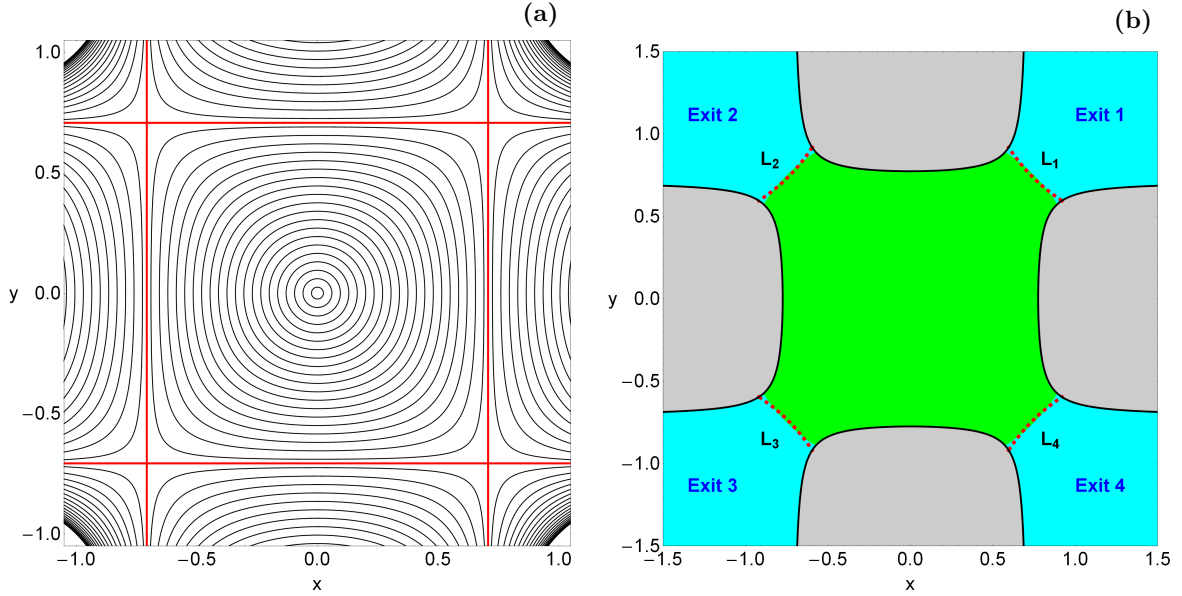}}
\caption{(a-left): Equipotential curves of the potential (\ref{pot}) for various values of the energy $h$. The equipotential curve corresponding to the energy of escape is shown with red color; (b-right): The open ZVC at the configuration $(x,y)$ plane when $h = 0.30$. $L_i$,  $i = 1, ..., 4$ indicate the four unstable Lyapunov orbits plotted in red. The interior region is indicated in green, the exterior region is shown in cyan, while the forbidden regions of motion are marked with grey.}
\label{cont}
\end{figure*}

Potential (\ref{pot}) has a finite energy of escape $(h_{esc})$ which is equal to 1/4. The equipotential curves of the potential (\ref{pot}) for various values of the energy $h$ are shown in Fig. \ref{cont}a. The equipotential corresponding to the energy of escape $(h_{esc})$ is plotted with red color in the same plot. The open Zero Velocity Curve (ZVC) at the configuration $(x,y)$ plane when $h = 0.30 > h_{esc}$ is presented with black color in Fig. \ref{cont}b and the four channels of escape are shown. In the same plot we denote the four unstable Lyapunov orbits by $L_i$, $i = 1, ..., 4$ using red color. We may say that the four exits act as hoses connecting the interior region (green color) with the ``outside world" of the exterior region (cyan color), while the forbidden regions of motion are shown with gray.

\section{Computational methods}
\label{cometh}

In order to explore the orbital properties in our Hamiltonian system we need to define samples of orbits whose nature will be identified. For this purpose we define for each value of the energy (all tested energy levels are above the escape energy), dense uniform grids of $1024 \times 1024$ initial conditions regularly distributed in the area allowed by the value of the energy. Our numerical investigation takes place in both the configuration $(x,y)$ and the phase $(x,\dot{x})$ space for a better understanding of the escape mechanism. For each initial condition, we integrated the equations of motion (\ref{eqmot}) as well as the variational equations (\ref{variac}) using a double precision Bulirsch-Stoer \verb!FORTRAN 77! algorithm (e.g., [\citealp{PTVF92}]) with a small time step of order of $10^{-2}$, which is sufficient enough for the desired accuracy of our computations (i.e., our results practically do not change by halving the time step). Our previous experience suggests that the Bulirsch-Stoer integrator is both faster and more accurate than a double precision Runge-Kutta-Fehlberg algorithm of order 7 with Cash-Karp coefficients. In all cases, the energy integral (Eq. (\ref{ham})) was conserved better than one part in $10^{-11}$, although for most orbits it was better than one part in $10^{-12}$.

An issue of paramount importance is the determination of the position as well as the time at which an orbit escapes. When the value of the energy $h$ is smaller than the escape energy, the ZVC is closed. On the other hand, when $h > h_{esc}$ the equipotential curves are open and extend to infinity. An open ZVC consists of several branches forming channels through which an orbit can escape to infinity. At every opening there is a highly unstable periodic orbit close to the line of maximum potential [\citealp{C79}] which is called a Lyapunov orbit. Such an orbit reaches the ZVC on both sides of the opening and returns along the same path thus connecting two opposite branches of the ZVC. Lyapunov orbits are very important for the escapes from a Hamiltonian system, since if an orbit intersects any one of these orbits with velocity pointing outwards moves always outwards and eventually escapes from the system without any further intersections with the surface of section (see e.g., [\citealp{C90}]). The passage of orbits through Lyapunov orbits and their subsequent escape to infinity is the most conspicuous aspect of the transport, but crucial features of the bulk flow, especially at late times, appear to be controlled by diffusion through cantori, which can trap orbits far vary long time periods.

In our computations, we set $10^5$ time units as a maximum time of numerical integration. Our previous experience in this subject indicates, that usually orbits need considerable less time to find one of the exits in the limiting curve and eventually escape from the system (obviously, the numerical integration is effectively ended when an orbit passes through one of the escape channels and intersects one of the unstable Lyapunov orbits). Nevertheless, we decided to use such a vast integration time just to be sure that all orbits have enough time in order to escape. Remember, that there are the so called ``sticky orbits" which behave as regular ones and their true chaotic character is revealed only after long time intervals of numerical integration. Here we should clarify, that orbits which do not escape after a numerical integration of $10^5$ time units are considered as non-escaping or trapped.

The configuration and the phase space are divided into the escaping and non-escaping or trapped domains. Usually, the vast majority of the non-escaping space is occupied by initial conditions of regular orbits forming stability islands where a third integral is present. In many systems however, trapped chaotic orbits have also been observed. Therefore, we decided to distinguish between regular and chaotic trapped orbits. Over the years, several chaos indicators have been developed in order to determine the character of orbits. In our case, we chose to use the Smaller ALingment Index (SALI) method. The SALI [\citealp{S01}] has been proved a very fast, reliable and effective tool, which is defined as
\begin{equation}
\rm SALI(t) \equiv min(d_-, d_+),
\label{sali}
\end{equation}
where $d_- \equiv \| {\bf{w_1}}(t) - {\bf{w_2}}(t) \|$ and $d_+ \equiv \| {\bf{w_1}}(t) + {\bf{w_2}}(t) \|$ are the alignments indices, while ${\bf{w_1}}(t)$ and ${\bf{w_2}}(t)$, are two deviations vectors which initially point in two random directions. For distinguishing between ordered and chaotic motion, all we have to do is to compute the SALI along time interval $t_{max}$ of numerical integration. In particular, we track simultaneously the time-evolution of the main orbit itself as well as the two deviation vectors ${\bf{w_1}}(t)$ and ${\bf{w_2}}(t)$ in order to compute the SALI. The variational equations (\ref{variac}), as usual, are used for the evolution and computation of the deviation vectors.

The time-evolution of SALI strongly depends on the nature of the computed orbit since when the orbit is regular the SALI exhibits small fluctuations around non zero values, while on the other hand, in the case of chaotic orbits the SALI after a small transient period it tends exponentially to zero approaching the limit of the accuracy of the computer $(10^{-16})$. Therefore, the particular time-evolution of the SALI allow us to distinguish fast and safely between regular and chaotic motion (e.g., [\citealp{ZC13}]). Nevertheless, we have to define a specific numerical threshold value for determining the transition from regularity to chaos. After conducting extensive numerical experiments, integrating many sets of orbits, we conclude that a safe threshold value for the SALI is the value $10^{-7}$. In order to decide whether an orbit is regular or chaotic, one may use the usual method according to which we check after a certain and predefined time interval of numerical integration, if the value of SALI has become less than the established threshold value. Therefore, if SALI $\leq 10^{-7}$ the orbit is chaotic, while if SALI $ > 10^{-7}$ the orbit is regular. For the computation of SALI we used the \verb!LP-VI! code [\citealp{CMD14}], a fully operational code which efficiently computes a suite of many chaos indicators for dynamical systems in any number of dimensions.

\section{Numerical results}
\label{numres}

The main objective is to distinguish between trapped, escaping and non-escaping orbits for values of energy larger than the escape energy where the ZVC is open and four channels of escape are present. In addition, two important properties of the orbits will be examined: (i) the directions or channels through which the particles escape and (ii) the time-scale of the escapes (we shall also use the term escape period).

\subsection{The configuration $(x,y)$ space}
\label{cas1}

For a constant value of the energy $h = const$ the motion is restricted to a three-dimensional surface. With polar coordinates $(r, \phi)$ in the center (0,0) the condition $\dot{r} = 0$ defines a two-dimensional surface od section in the surface $h = const$, with two disjoint parts $\dot{\phi} < 0$ and $\dot{\phi} > 0$. Each of these two parts has a unique projection onto the configuration $(x,y)$ plane. We decided to explore the escape process in the $\dot{\phi} > 0$ part of the configuration $(x,y)$ plane (the $\dot{\phi} < 0$ part gives very similar results).

\begin{figure*}[!tH]
\centering
\resizebox{0.90\hsize}{!}{\includegraphics{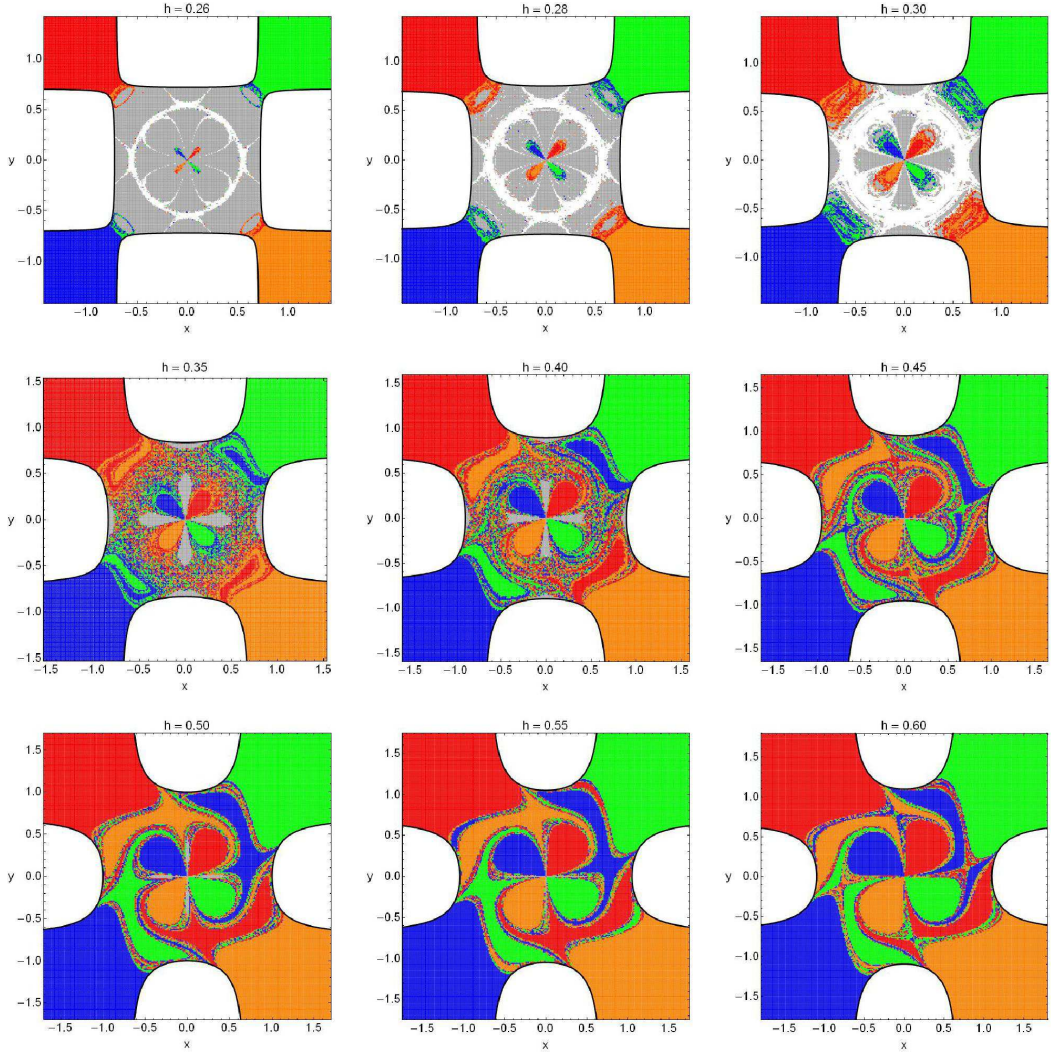}}
\caption{The structure of the configuration $(x,y)$ plane for several values of the energy $h$, distinguishing between different escape channels. The color code is as follows: Non-escaping regular (gray); trapped chaotic (white); escape through channel 1 (green); escape through channel 2 (red); escape through channel 3 (blue); escape through channel 4 (orange).}
\label{xy}
\end{figure*}

\begin{figure}[!tH]
\centering
\includegraphics[width=0.5\hsize]{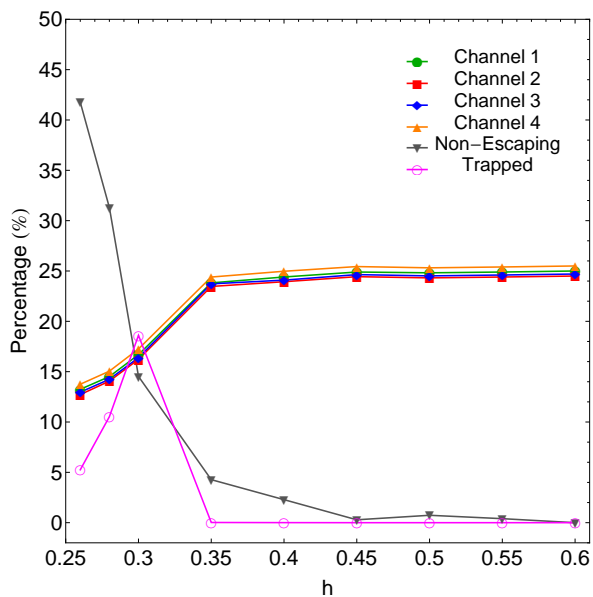}
\caption{Evolution of the percentages of trapped, escaping and non-escaping orbits on the configuration $(x,y)$ space when varying the energy $h$.}
\label{percs1}
\end{figure}

\begin{figure*}[!tH]
\centering
\resizebox{0.95\hsize}{!}{\includegraphics{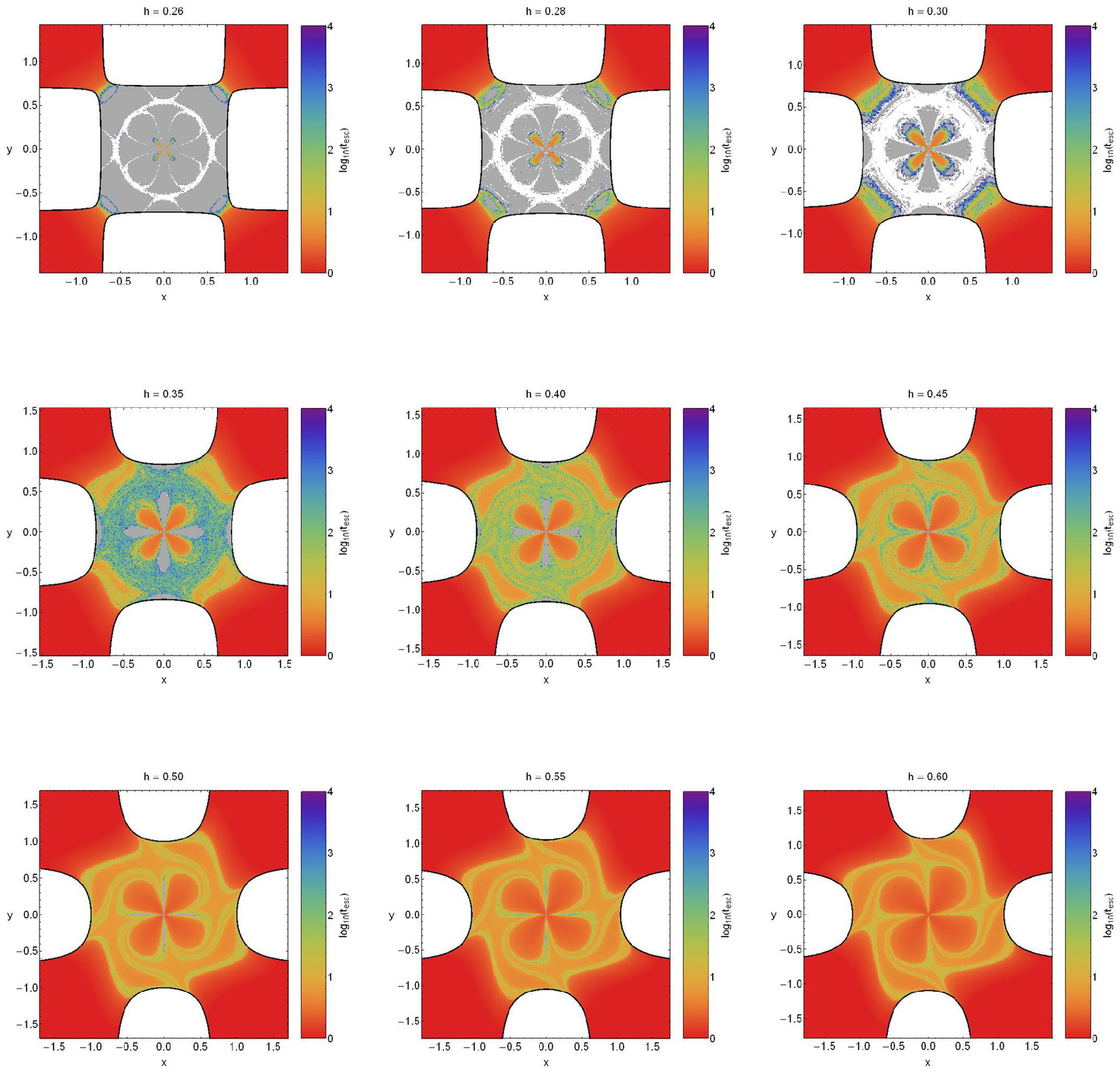}}
\caption{Distribution of the escape times $t_{\rm esc}$ of the orbits on the configuration $(x,y)$ space. The darker the color, the larger the escape time. Non-escaping regular orbits are indicated by gray color, while trapped chaotic orbits are shown in white.}
\label{txy}
\end{figure*}

In our investigation, we shall deal only with unbounded motion of test particles for values of energy in the set $h = \{0.26, 0.28, 0.30, 0.35, 0.40, 0.45, 0.50, 0.55, 0.60\}$. Fig. \ref{xy} shows the structure of the configuration $(x,y)$ plane for different values of the energy. Each initial condition is colored according to the escape channel through which the particular orbit escapes. The gray regions on the other hand, denote initial conditions where the test particles move in regular orbits and therefore do not escape, while white dots correspond to trapped chaotic orbits. The outermost black solid line is the ZVC (or limiting curve) which is defined as $V(x,y) = h$. It is seen that for $h = 0.26$, that is an energy level just above the escape energy, almost all the interior region is covered by trapped or non-escaping orbits. In particular, we observe that initial conditions corresponding to trapped chaotic orbits form ring-shaped structures. Moreover, near the center of the interior region we identify a propeller-shaped structure corresponding to escaping orbits. As we proceed to higher energy levels (up to $h = 0.30$) the ring structures increase, while the non-escaping domains decrease. For $h = 0.35$, trapped motion has almost completely disappeared giving its place to a highly fractal domain. Furthermore, the non-escaping orbits are located either near the boundaries of the ZVC or inside a cross-shaped area near the center of the interior region. As the value of the energy increases more we observe three main alterations regarding the orbital structure of the configuration space: (i) the propeller-shaped region increases, (ii) the cross-shaped region decreases and (iii) the intermediate area becomes less and less fractal. At the highest energy level studied $(h = 0.60)$ there is no indication of non-escaping regular motion and all integrated orbits escape to infinity through one of the four escape channels. It is also seen that almost all the configuration space is dominated by well formed basins of escape, while only the domains between the several escape basins remain fractal.

It is of particular interest to monitor the evolution of the percentages of trapped, escaping and non-escaping orbits on the configuration $(x,y)$ plane when the value of the energy $h$ varies. A diagram depicting this evolution is presented in Fig. \ref{percs1}. We see, that for $h = 0.26$ about 42\% of the configuration space is covered by initial conditions of non-escaping regular orbits. As the value of the energy increases however, the rate of non-escaping regular orbits drops rapidly and eventually at $h = 0.6$ it vanishes. Initially the rate of trapped chaotic orbits is about 5\% and for $0.26 < h < 0.30$ it exhibits a sudden increase. For $h > 0.30$ however, it reduces and for $h > 0.35$ is effectively zero. We also observe that the evolution of the percentages of the escaping orbits display a common behaviour due to the symmetry of the configuration space. In particular, for $0.26 < h < 0.35$ the percentages increase, while for $h > 0.35$ they remain almost unperturbed at about 25\% thus implying that all escape channels are equiprobable.

The following Fig. \ref{txy} shows how the escape times $t_{\rm esc}$ of orbits are distributed on the $(x,y)$ space. Light reddish colors correspond to fast escaping orbits, dark blue/purple colors indicate large escape periods, gray color denote non-escaping regular orbits, while white areas correspond to trapped chaotic orbits. Here, we have a better view regarding the amount of non-escaping and trapped orbits. We observe, that when $h = 0.26$, that is a value of energy very close to the escape energy, the escape periods of the majority of orbits in the boundaries between the interior and exterior regions are huge corresponding to tens of thousands of time units. This however, is anticipated because in this case the width of the escape channels is very small and therefore, the orbits should spend much time inside the equipotential curve until they find one of the four openings and eventually escape to infinity. As the value of the energy increases however, the escape channels become more and more wide leading to faster escaping orbits, which means that the escape period decreases rapidly. We found, that the longest escape rates correspond to initial conditions near the boundaries between the escape basins and near the vicinity of stability islands. On the other hand, the shortest escape periods have been measured for the regions without sensitive dependence on the initial conditions (basins of escape), that is, those far away from the fractal basin boundaries.

\subsection{The phase $(x,\dot{x})$ space}
\label{cas2}

\begin{figure*}[!tH]
\centering
\resizebox{0.90\hsize}{!}{\includegraphics{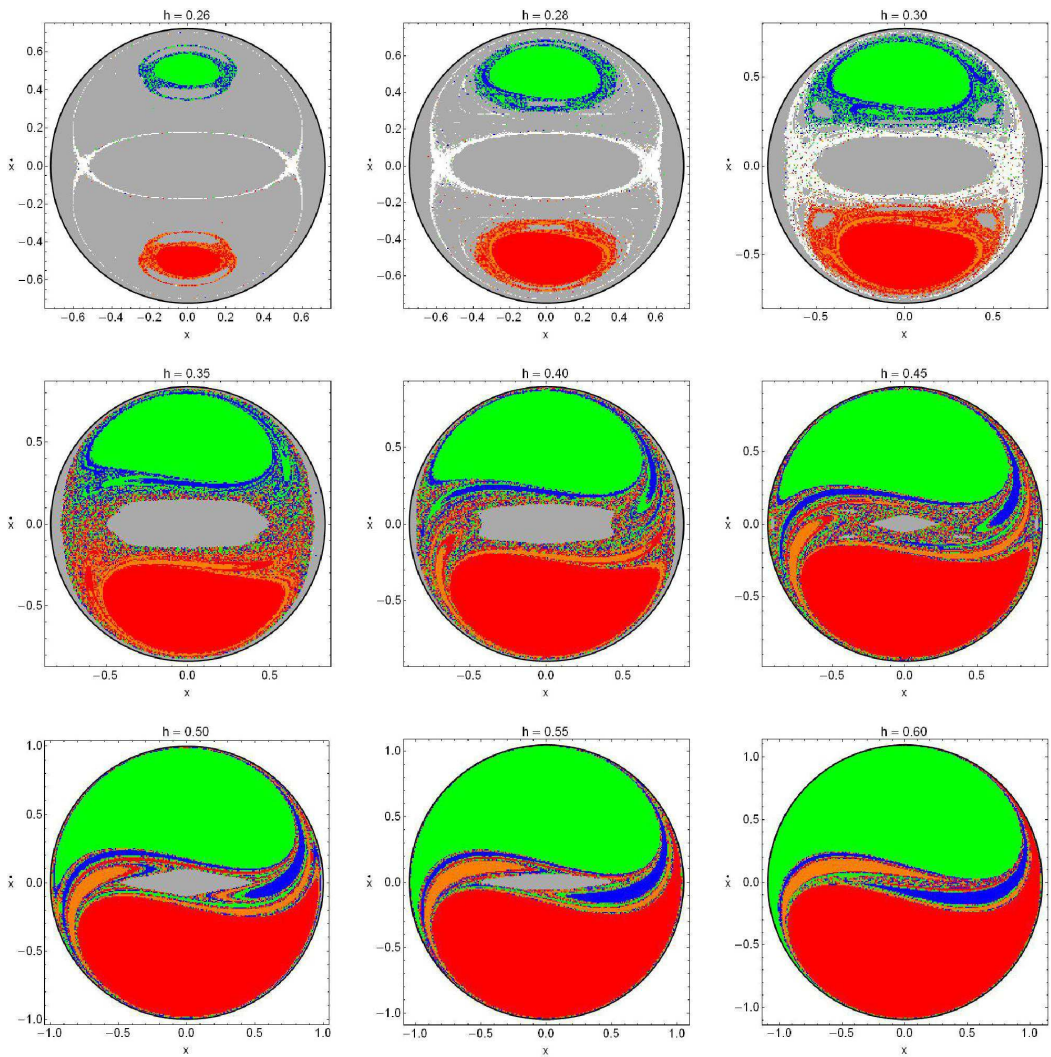}}
\caption{The structure of the phase $(x,\dot{x})$ plane for several values of the energy $h$, distinguishing between different escape channels. The color code is as follows: Non-escaping regular (gray); trapped chaotic (white); escape through channel 1 (green); escape through channel 2 (red); escape through channel 3 (blue); escape through channel 4 (orange).}
\label{xpx}
\end{figure*}

\begin{figure}[!tH]
\centering
\includegraphics[width=0.5\hsize]{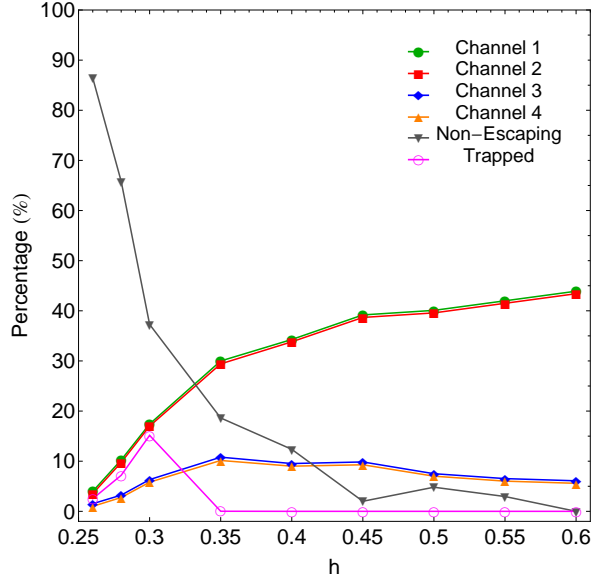}
\caption{Evolution of the percentages of trapped, escaping and non-escaping orbits on the phase $(x,\dot{x})$ space when varying the energy $h$.}
\label{percs2}
\end{figure}

\begin{figure*}[!tH]
\centering
\resizebox{0.95\hsize}{!}{\includegraphics{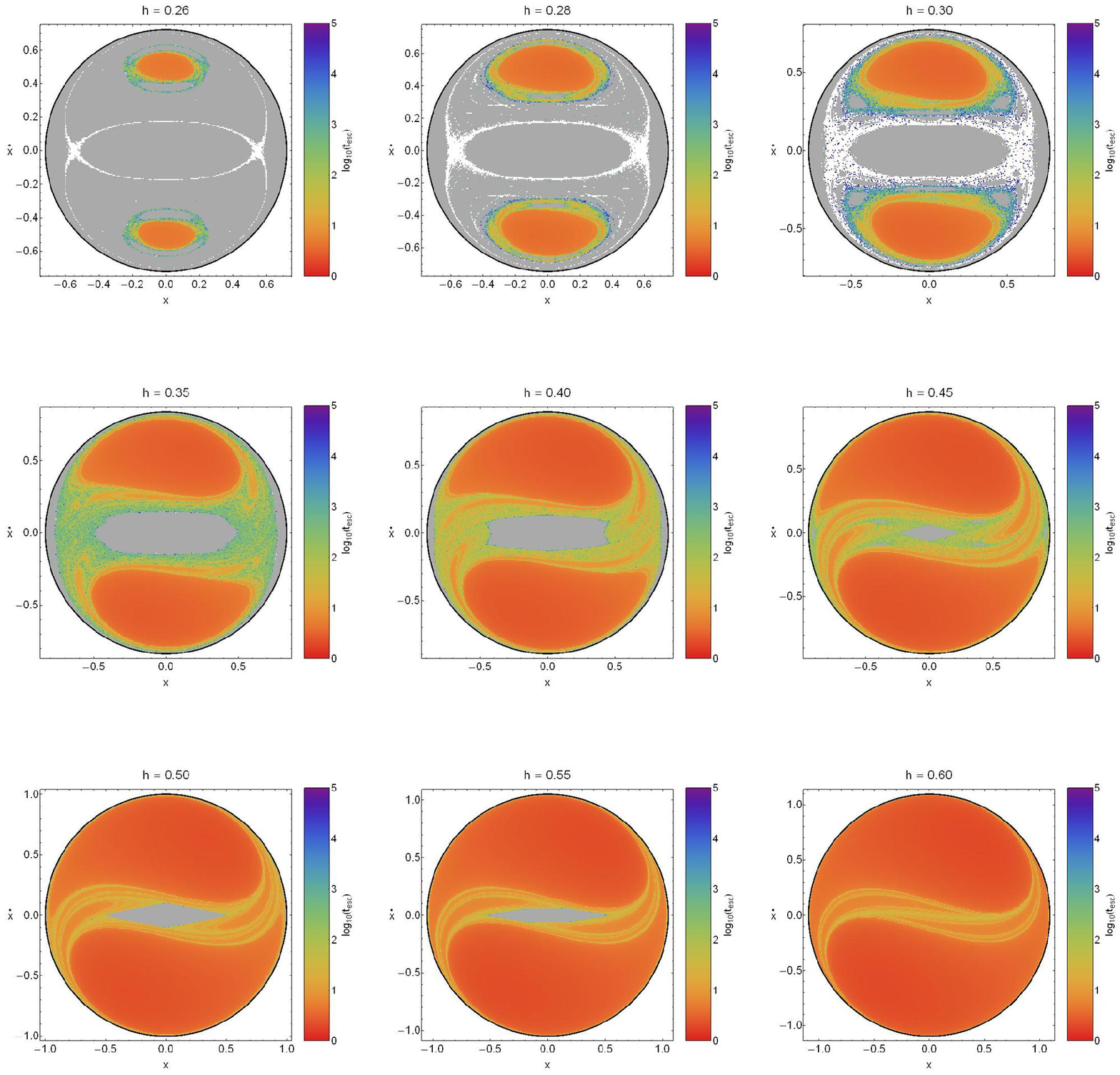}}
\caption{Distribution of the escape times $t_{\rm esc}$ of the orbits on the phase $(x,\dot{x})$ space. The darker the color, the larger the escape time. Non-escaping regular orbits are indicated by gray color, while trapped chaotic orbits are shown in white.}
\label{txpx}
\end{figure*}

We continue our exploration of the escape process in the phase $(x,\dot{x})$ space. The structure of the $(x,\dot{x})$ phase plane for the same set of values of the energy is shown in Fig. \ref{xpx}. We observe a similar behavior to that discussed for the configuration $(x,y)$ plane in Fig. \ref{xy}. The outermost black solid line is the limiting curve which is defined as
\begin{equation}
f(x,\dot{x}) = \frac{1}{2}\dot{x}^2 + V(x, y = 0) = h.
\label{zvc}
\end{equation}
It is worth noticing, that in the phase plane the limiting curve is closed but this does not mean that there is no escape. Remember, that we decided to choose such perturbation terms that produce the escape channels on the configuration $(x,y)$ plane which is a subspace of the entire four-dimensional $(x,y,\dot{x},\dot{y})$ space of the system. Here we must point out, that this $(x,\dot{x})$ phase plane is not a Poincar\'{e} Surface of Section (PSS), simply because escaping orbits in general, do not intersect the $y = 0$ axis after a certain time, thus preventing us from defying a recurrent time. A classical Poincar\'{e} surface of section exists only if orbits intersect an axis like $y = 0$ at least once within a certain time interval. Nevertheless, in the case of escaping orbits we can still define local surfaces of section which help us to understand the orbital behavior of the dynamical system.

In this case, the grids of initial conditions of orbits whose properties will be examined are defined as follows:  we consider orbits with initial conditions $(x_0, \dot{x_0})$ with $y_0 = 0$, while the initial value of $\dot{y_0}$ is always obtained from the energy integral (\ref{ham}) as $\dot{y_0} = \dot{y}(x_0,y_0,\dot{x_0},h) > 0$.

One may observe, that for $h < 0.3$ the majority of the phase plane is covered by a vast region corresponding to non-escaping regular orbits, while only two small islands of initial conditions of escaping orbits are present. It is interesting to note the existence of a thin separatrix composed of trapped chaotic orbits. However, as the value of the energy increases and we move away for the escape energy, the extent of these two islands grows and for $h > 0.35$ the non-escaping regular orbits are mainly confined to the central region of the phase plane. At the same time, small elongated spiral basins of escape emerge inside the fractal region which surrounds the area of non-escaping orbits. Furthermore, at very high energy levels $(h > 0.55)$ we see that trapped orbits disappear completely from the grid and the two main basins of escape take over the vast majority of the phase plane, while the elongated escape basins remain confined to the central region. As we noticed previously when discussing the configuration $(x,y)$ plane, there is also a symmetry in the phase plane. In particular, throughout the energy range the structure of the phase plane $(x,\dot{x})$ is somehow symmetrical (not with the strick sense) with respect to the $\dot{x} = 0$ axis.

The evolution of the percentages of trapped and escaping orbits on the phase plane as a function of the value of the energy $h$ is given in Fig. \ref{percs2}. For $h = 0.26$, we see that non-escaping regular orbits dominate the phase plane as they occupy about 90\% of the gird. However as usual, with increasing energy the dominance of non-escaping orbits deteriorates rapidly due to the increase of the rates of escaping orbits which form basins of escape. Moreover, trapped chaotic orbits are present only for relatively small energy levels $(h < 0.30)$. We observe that the evolution of the percentages of orbits escaping through channels 1 and 2 coincides with the evolution of the percentages escaping trough channels 3 and 4, respectively. The percentages of all types of escaping orbits increase but with different rates and for $h > 0.35$ they overwhelm the amount of non-escaping regular orbits. In particular, we see that the percentages of orbits escaping through exits 1 and 2 are always higher than those corresponding to orbits escaping through channels 3 and 4. Moreover, the rates of exits 1 and 2 increase constantly and at the highest energy level studied $(h = 0.6)$ they share about 90\% of the entire phase plane. On the other hand, the percentages of exits 3 and 4, even though they also grow with increasing energy, they always possess significantly smaller values than exits 1 and 2 and for $h > 0.4$ they seem to saturate around 5\%. Thus, we may conclude that the vast majority of orbits in the phase $(x,\dot{x})$ space exhibit clear sings of preference through exits 1 and 2, while channels 3 and 4 have considerable less probability to be chosen.

The following Fig. \ref{txpx} shows the distribution of the escape times $t_{\rm esc}$ of orbits on the $(x,\dot{x})$ space. It is evident that orbits with initial conditions inside the exit basins escape to infinity after short time intervals, or in other words, they possess extremely small escape periods. On the contrary, orbits with initial conditions located in the fractal parts of the phase plane need considerable amount of time in order to find one of the four exits and escape. It is seen that at the highest energy level studied $(h = 0.6)$ there is no indication of bounded motion and all orbits escape to infinity sooner or later.

\subsection{An overview analysis}
\label{geno}

\begin{figure*}[!tH]
\centering
\resizebox{\hsize}{!}{\includegraphics{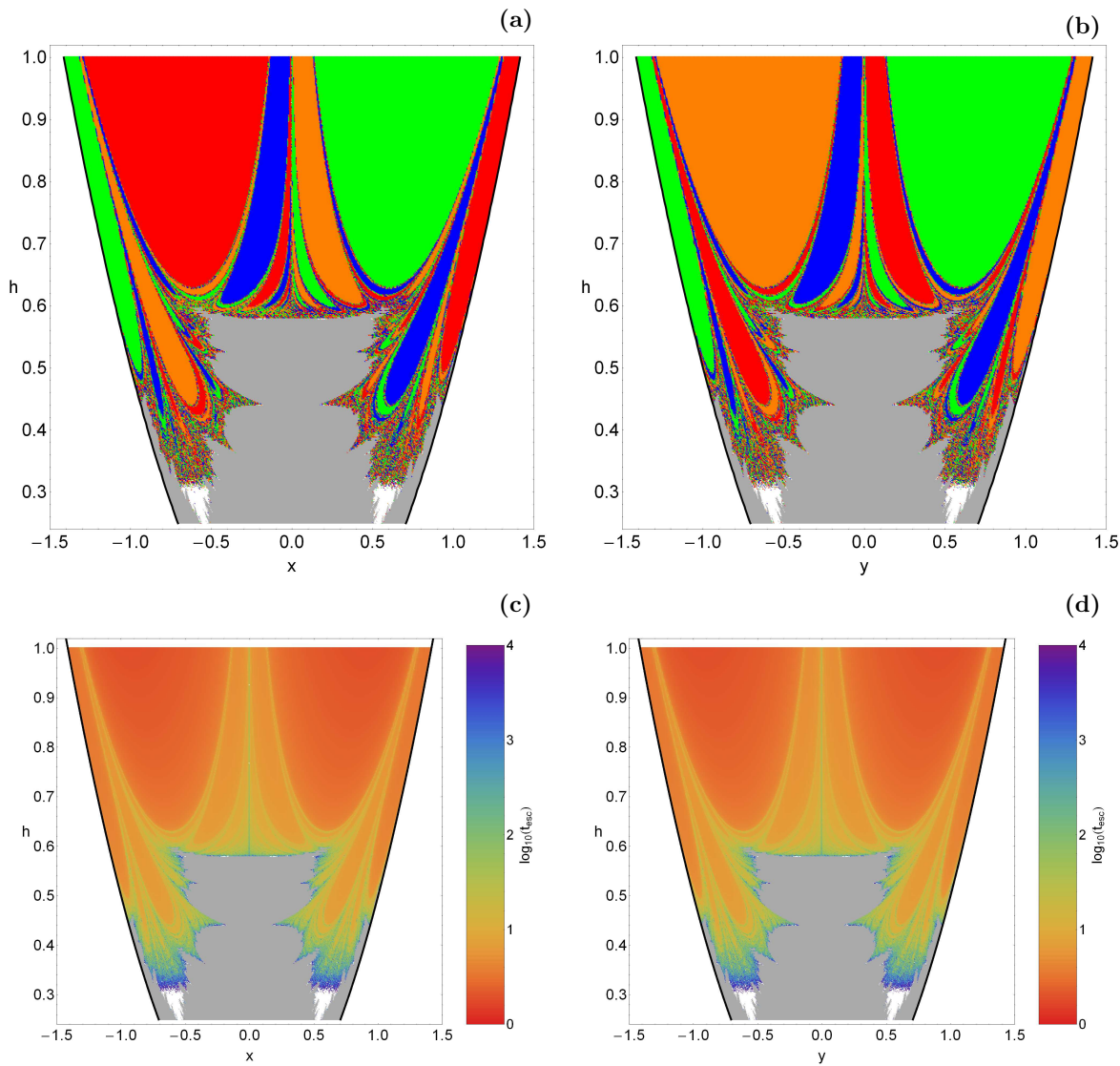}}
\caption{Orbital structure of the (a-left): $(x,h)$-plane and (b-upper right): $(y,h)$-plane. These diagrams provide a detailed analysis of the evolution of the trapped, escaping and non-escaping orbits of the Hamiltonian. The color code is the same as in Fig. \ref{xy}. (c-d): The distribution of the corresponding escape times of the orbits. In this type of grid representation the stability islands of regular non-escaping orbits which are indicated by gray color and the trapped chaotic orbits shown in white can be identified more easily.}
\label{xyht}
\end{figure*}

The color-coded grids in configuration $(x,y)$ as well as the phase $(x,\dot{x})$ plane provide information on the phase space mixing however, for only a fixed value of energy. H\'{e}non back in the late 60s [\citealp{H69}], introduced a new type of plane which can provide information not only about stability and chaotic regions but also about areas of trapped and escaping orbits using the section $y = \dot{x} = 0$, $\dot{y} > 0$ (see also [\citealp{BBS08}]). In other words, all the orbits of the test particles are launched from the $x$-axis with $x = x_0$, parallel to the $y$-axis $(y = 0)$. Consequently, in contrast to the previously discussed types of planes, only orbits with pericenters on the $x$-axis are included and therefore, the value of the energy $h$ can be used as an ordinate. In this way, we can monitor how the energy influences the overall orbital structure of our Hamiltonian system using a continuous spectrum of energy values rather than few discrete energy levels. Fig. \ref{xyht}a shows the structure of the $(x,h)$-plane when $h \in (0.25, 1]$. In order to obtain a more complete view of the orbital structure of the system, we follow a similar numerical approach to that explained before but in this case we use the section $x = \dot{y} = 0$, $\dot{x} > 0$, considering orbits that are launched from the $y$-axis with $y = y_0$, parallel to the $x$-axis. This allow us to construct again a two-dimensional (2D) plane in which the $y$ coordinate of orbits is the abscissa, while the value of the energy $h$ is the ordinate. Fig \ref{xyht}b shows the structure of the $(y,h)$ plane, while the distribution of the corresponding escape time of orbits is given in Fig. \ref{xyht}(c-d).

\begin{figure*}[!tH]
\centering
\resizebox{\hsize}{!}{\includegraphics{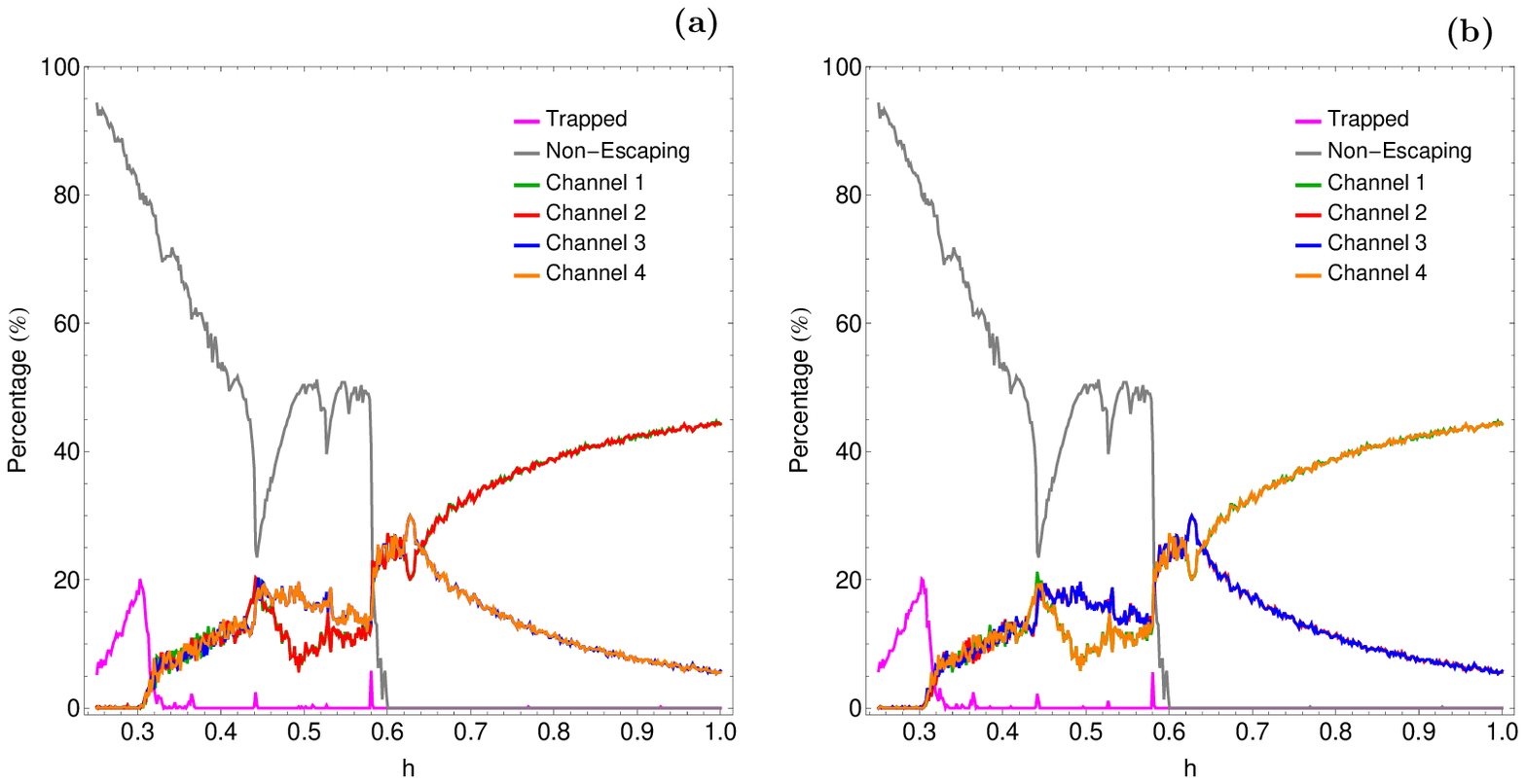}}
\caption{Evolution of the percentages of trapped, escaping and non-escaping orbits on the (a-left): $(x,h)$-plane and (b-right): $(y,h)$-plane, as a function of the energy $h$.}
\label{percs3}
\end{figure*}

It is seen in both types of planes that the boundaries between bounded and unbounded motion are now seen to be more jagged than in the previous types of grids. In addition, we found in the blow-ups of the diagrams many tiny islands of stability\footnote{From chaos theory we expect an infinite number of islands of (stable) quasi-periodic (or small scale chaotic) motion.}. We observe that for low values of the energy close to the escape energy, there is a considerable amount of non-escaping orbits inside stability regions surrounded by a highly fractal structure. This pattern however changes for larger energy levels $(h > 0.60)$, where there are no non-escaping regular orbits and the vast majority of the grids is covered by well-formed basins of escape, while fractal structure is confined only near the boundaries of the escape basins. Finally, it is interesting to note the presence of trapped chaotic orbits for very low energy levels $(h < 0.30)$. The overall structure of the $(y,h)$-plane is identical to that of the $(x,h)$-plane however there is major difference: the escape basins corresponding to exit channels 2 and 4 change places with each other.

In Fig. \ref{percs3}(a-b) we present the evolution of the percentages of all types of orbits as a function of the orbital energy $h$ for both types of planes. We observe that just above the escape energy about 95\% of the planes is covered by initial conditions corresponding to non-escaping regular orbits. As the value of the energy increases however, the rate of non-escaping orbits drops and finally when $h > 0.6$ it vanishes. Trapped chaotic orbits on the other hand, are observed mainly at low energy levels (occupying about 20\% of the planes when $h = 0.30$), while for $h > 0.60$ there is no indication of trapped chaos. The percentages of escaping orbits evolve almost identically up to $h = 0.64$, while for larger values of energy they start to diverge. In particular it is seen that the rates of exit channels 1 and 2 (for the $(x,h)$-plane) and 1 and 4 (for the $(y,h)$-plane) increase, while on the other hand, the percentages of the other two exit channels decrease. At the highest energy level studied $(h = 1.0)$ escaping orbits trough the first couples of exit channels share about 90\% of the plane, while the other two exit channels cover only about 10\% of the same planes.

\begin{figure}[!tH]
\centering
\includegraphics[width=0.5\hsize]{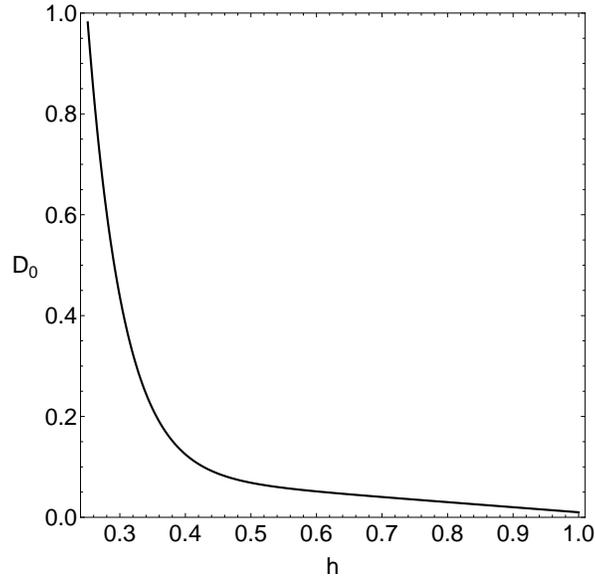}
\caption{Evolution of the fractal dimension $D_0$ of the $(x,h)$-plane shown in Figs. \ref{xyht}a as a function of the total energy $h$. $D_0 = 1$ means total fractality, while $D = 0$ implies zero fractality.}
\label{frac}
\end{figure}

In the previous two subsections we discuss the fractality of the phase space in a rather qualitative way. In particular, rich and highly fractal domains are those in which we cannot predict through which exit channel the particle will escape since the particle chooses randomly an exit. On the other hand, inside the escape basins where the degree of fractality is zero the escape process of the particles is well known and predictable. At this point, we shall provide a quantitative analysis of the degree of fractality for the grids shown in Figs. \ref{xyht}(a-b). In order to measure the fractality we have computed the uncertainty dimension [\citealp{O93}] for different values of the total energy following the computational method introduced in [\citealp{AVS01}]. Obviously, this quantity is independent of the initial conditions used to compute it. The way to do it is the following. We calculate the exit for a certain initial condition $(x,h)$ and $(y,h)$. Then, we compute the exit for the initial conditions $(x - \epsilon, h)$, $(x + \epsilon, h)$ and $(y - \epsilon, h)$, $(y + \epsilon, h)$ for a small $\epsilon$ and if all of them coincide, then this point is labeled as ``certain". If on the other hand they do not, it will be labeled as ``uncertain". We repeat this procedure for different values of $\epsilon$. Then we calculate the fraction of initial conditions that lead to uncertain final states $f(\epsilon)$. There exists a power law between $f(\epsilon)$ and $\epsilon$, $f(\epsilon) \propto \epsilon^{\alpha}$, where $\alpha$ is the uncertainty exponent. The uncertainty dimension $D_0$ of the fractal set embedded in the initial conditions is obtained from the relation $D_0 = D - \alpha$, where $D$ is the dimension of the phase space. It is typical to use a fine grid of values of $x$ or $y$ and $h$ to calculate the uncertainty dimension. The evolution of the uncertainty dimension $D_0$ of the $(x,h)$ plane as a function of the energy is shown in Fig. \ref{frac} (For the $(y,h)$ plane the corresponding diagram is completely the same due to the symmetry of the two planes). As it has just been explained, the computation of the uncertainty dimension is done for only a ``1D slice'' of initial conditions of Figs. \ref{xyht}(a-b), and for that reason $D_0 \in (0,1)$. It is remarkable that the uncertainty dimension tends to one when the energy tends to its minimum value $E_{esc} = 1/4$. This means that for that critical value, there is a total fractalization of the phase space, and the chaotic set becomes ``dense" in the limit. Consequently, in this limit there are no smooth sets of initial conditions (see Figs. \ref{xyht}) and the only defined structures that can be recognized are the Kolmogorov-Arnold-Moser (KAM)-tori of quasi-periodic orbits. When the energy is increased however, the different smooth sets appear and tend to grow, while the fractal structures that coincide with the boundary between basins decrease. Finally for values of energy much greater than the escape energy the uncertainty dimension tends to zero (no fractality).

The rich fractal structure of the $(x,h)$ and $(y,h)$ planes observed in Figs. \ref{xyht}(a-b) implies that the system has also a strong topological property, which is known as the Wada property [\citealp{AVS01}]. The Wada property is a general feature of two-dimensional (2D) Hamiltonians with three or more escape channels. A basin of escape verifies the property of Wada if any initial condition that is on the boundary of one basin is also simultaneously on the boundary of three or even more escape basins (e.g., [\citealp{BSBS12}, \citealp{KY91}]). In other words, every open neighborhood of a point $x$ belonging to a Wada basin boundary has a nonempty intersection with at least three different basins. Hence, if the initial conditions of a particle are in the vicinity of the Wada basin boundary, we will not be able to be sure by which one of the three exits the orbit will escape to infinity. Therefore, if a Hamiltonian system verifies the property of Wada, the unpredictability is even stronger than if it only had fractal basin boundaries [\citealp{BGOB88}, \citealp{O93}]. If an orbit starts close to any point in the boundary, it will not be possible to predict its future behavior, as its initial conditions could belong to any of the other escape basins. This special topological property has been identified and studied in several dynamical systems (e.g., [\citealp{AVS09}, \citealp{KY91}, \citealp{PCOG96}]) and it is a typical property in open Hamiltonian systems with three or more escape channels.

\section{Conclusions and discussion}
\label{disc}

In this work we numerically investigated the escape properties of orbits in a Hamiltonian system of two-dimensional coupled perturbed harmonic oscillators with four exit channels, which is a characteristic example of open Hamiltonian systems. The key feature of this type of Hamiltonians is that they have a finite energy of escape. In particular, for energies smaller than the escape value, the equipotential surfaces are close and therefore escape is impossible. For energy levels larger than the escape energy however, the equipotential surfaces open and several channels of escape appear through which the particles can escape to infinity. Here we should emphasize, that if a test particle has energy larger than the escape value, this does not necessarily mean that the particle will certainly escape from the system and even if escape does occur, the time required for an orbit to cross a Lyapunov orbit and hence escape to infinity may be vary long compared with the natural crossing time. The function containing the perturbation terms affects significantly the structure of the equipotential curves and determines the exact number of the escape channels.

We defined for each value of the energy, dense uniform grids of $1024 \times 1024$ initial conditions regularly distributed in the area allowed by the value of the energy in both the configuration and the phase space. For the numerical integration of the orbits in each grid, we needed roughly between 1 minute and 3.5 days of CPU time on a Pentium Dual-Core 2.2 GHz PC, depending both on the amount of bounded (trapped and non-escaping) orbits and on the escape rates of orbits in each case. For each initial condition, the maximum time of the numerical integration was set to be equal to $10^5$ time units however, when a particle escapes the numerical integration is effectively ended and proceeds to the next initial condition.

By conducting a thorough and systematical numerical investigation we successfully revealed the structure of both the configuration and the phase space. In particular, we managed to distinguish between trapped (non-escaping) and escaping orbits and we located the basins of escape leading to different exit channels, also finding correlations with the corresponding escape times of the orbits. Among the escaping orbits, we separated between those escaping fast or late from the system. Our extensive numerical calculations strongly suggest, that the overall escape process is very dependent on the value of the total orbital energy. The main numerical results of our investigation can be summarized as follows:
\begin{enumerate}
 \item Areas of trapped or non-escaping orbits and regions of initial conditions leading to escape in a given direction (basins of escape), were found to exist in both the configuration and the phase space. The several escape basins are very intricately interwoven and they appear either as well-defined broad regions or thin elongated spiral bands. Regions of trapped orbits first and foremost correspond to stability islands of regular orbits where a third integral of motion is present.
 \item A strong correlation between the extent of the basins of escape and the value of the energy $h$ was found to exists. Indeed, for low values of $h$ the structure of both configuration and phase space exhibits a large degree of fractalization and therefore the majority of orbits escape choosing randomly escape channels. As the value of $h$ increases however, the structure becomes less and less fractal and several basins of escape emerge. The extent of these basins of escape is more prominent at high energy levels, where they occupy about 90\% of the entire area on the grids.
 \item It was found, that for energy levels slightly above the escape energy the majority of the escaping orbits have considerable long escape rates (or escape periods), while as we proceed to higher energies the proportion of fast escaping orbits increases significantly. This phenomenon can be justified, if we take into account that with increasing energy the exit channels on the equipotential curves become more and more wide thus the test particles can find easily and faster one of the exits and escape to infinity.
 \item We observed, that in several exit regions the escape process is highly sensitive dependent on the initial conditions, which means that a minor change in the initial conditions of an orbit lead the test particle to escape through another exit channel. These regions are the opposite of the escape basins, are completely intertwined with respect to each other (fractal structure) and are mainly located in the vicinity of stability islands. This sensitivity towards slight changes in the initial conditions in the fractal regions implies, that it is impossible to predict through which exit the particle will escape.
 \item Our calculations revealed, that the escape times of orbits are directly linked to the basins of escape. In particular, inside the basins of escape as well as relatively away from the fractal domains, the shortest escape rates of the orbits had been measured. On the other hand, the longest escape periods correspond to initial conditions of orbits either near the boundaries between the escape basins or in the vicinity of stability islands.
 \item We found chaotic orbits with initial conditions close to the outermost KAM islands which remain trapped in the neighbourhood of these islands for vast time intervals having trapped periods which correspond to hundreds of thousands time units. Additional numerical integration shows that these chaotic orbits eventually escape.
 \item We provided numerical evidence that our Hamiltonian system has a strong topological property, known as the Wada property. This means that any initial condition that is on the boundary of an escape basin, is also simultaneously on the boundary of at leats other two basins of escape. We also concluded that if a dynamical system verifies the property of Wada, the unpredictability is even stronger than if it only had fractal basin boundaries.
\end{enumerate}

We hope that the present numerical results to be useful in the active field of open Hamiltonian systems which may have implications in different aspects of chaotic scattering with applications in several areas of physics. Furthermore, it is also in our future plans to expand our exploration in other more complicated potentials, focusing our interest in reveling the escape process of stars in galactic systems such as open star clusters, or binary stellar systems of interacting galaxies.

\section*{Acknowledgments}

The author would like to express his warmest thanks to the anonymous referees for the careful reading of the manuscript and for all the apt suggestions and comments which allowed us to improve both the quality and the clarity of the paper.

\label{pagefin}

\end{document}